# Invariant ionic conductance in an atomically thin polar nanopore

Shengping Zhang[1,2#], Haiou Zeng[1#], Ningran Wu[1,2#], Guodong Xue[2,3#], Xiao Li[1,2], Anshul Saxena[4], Junhe Tong[1], Nianjie Liang[1,5], Zeyu Zhuang[1], Jing Yang[1], Narayana R. Aluru[4], Kaihui Liu[3]*, Bai Song[1,5]*, and Luda Wang[1,2,6,7]*

[1]National Key Laboratory of Advanced Micro and Nano Manufacture Technology, School of Integrated Circuits, Peking University, Beijing 100871, China

[2]Academy for Advanced Interdisciplinary Studies and Center for Nanochemistry, Beijing Science and Engineering Center for Nanocarbons, Peking University, Beijing 100871, China

[3]State Key Laboratory for Mesoscopic Physics, School of Physics, Peking University, Beijing 100871, China

[4]Oden Institute for Computational Engineering and Sciences, Walker Department of Mechanical Engineering, The University of Texas at Austin, Austin, TX 78712, United States

[5]School of Mechanics and Engineering Science, Peking University, Beijing 100871, China

[6]Beijing Advanced Innovation Center for Integrated Circuits, Beijing 100871, China

[7]Beijing Graphene Institute, Beijing 100095, China

[#]These authors contributed equally to this work.

*Corresponding author. Email: luda.wang@pku.edu.cn (L.W.); songbai@pku.edu.cn (B.S.); khliu@pku.edu.cn (K.L.)

**ABSTRACT**

Ion channels regulate many essential properties of biological cells, especially the membrane potential. Despite decades of efforts on artificial channels, it remains a great challenge to mimic the dipole potential—an indispensable constituent of the membrane potential, due to its angstrom-scale characteristic length. Here, we explore nanopores in monolayer molybdenum sulfide selenide (MoSSe) considering its intrinsic dipole and atomic thickness. Remarkably, an invariant ionic conductance was observed over salt concentrations spanning six orders of magnitude, distinct from all known conductance-concentration scaling laws and reminiscent of the current saturation in cell membranes at high concentrations. Molecular dynamics simulations revealed the fundamental role of the dipole-modulated dielectric properties of nanoconfined water. Our findings highlight an exotic conductance scaling law and open up a novel avenue for controlling ion transport in unprecedented ways.

# I. INTRODUCTION

Transmembrane mass transport plays an indispensable role in various physiological activities including homeostasis and neurotransmission [1-4]. To precisely control the transport of ions and molecules, cell membranes have evolved sophisticated compositions, structures, and mechanisms. In particular, the membrane potential is of fundamental importance to the diverse functionalities of a cell, which consists of three kinds including the transmembrane potential, surface potential, and dipole potential [5,6]. The transmembrane potential originates from ion concentration differences across the lipid bilayer and is key to regulating voltage-gated ion channels. The surface potential arises from accumulated charges on the membrane surface and controls ion distribution near cell membranes [7]. The dipole potential is due to the aligned lipids and the arrangement of water molecules on the membrane surface[6]. In contrast to the transmembrane and surface potentials which are typically around 10 to 100 mV, the dipole potential can range between 100 and 1000 mV which is one to two orders of magnitude higher [8].

Inspired by various biological potentials, tremendous efforts have been devoted to fabricate artificial ion channels that can precisely modulate ion transport at the nanoscale and simulate or even surpass nature. For example, utilizing the transmembrane charge imbalance, salinity gradient has been widely explored in iontronic energy storage and conversion devices [9,10] and neuronal network activity modulation [11]. Moreover, surface potential has been employed to design the charge distribution along the transport pathway in order to regulate and rectify the flow of ions [12-15]. In contrast to the notable progress associated with the transmembrane and surface potentials, biomimetic studies of the dipole potential have received little attention impeding a thorough understanding of cell membranes and the imitation of their full functionality. Rational exploitation of the dipole potential is primarily limited by its characteristic length scale of only a few angstroms within the lipid bilayer. Further, decoupling the effect of the dipole potential presents an opportunity to understand transport mechanisms associated with dipole potentials.

In this work, we realize novel polar nanochannels that mimic the biological dipole potential and explore their intriguing ion transport properties. To begin with, we focus on monolayer two-dimensional (2D) materials which feature thicknesses comparable to the effective range of the dipole potential in cell membranes. In order to fabricate nanopores with a built-in dipole, it is natural to consider heterogeneous stacks. However, this at least doubles the membrane thickness. Instead, we employ a single layer of molybdenum sulfide selenide (MoSSe), an emerging Janus transition metal dichalcogenide (TMD) with different chalcogen atoms on its two opposite surfaces. This provides an intrinsic dipole inside the membrane with an ultimate length of 6.5 Å in the cross-plane direction [Fig. 1(a)], and enables us to investigate the impact of dipole potential on confined ion transport at the molecular level [Fig. 1(b)]. To this end, we focus on the aqueous solutions of potassium chloride.

Surprisingly, we observe a constant ionic conductance in MoSSe nanopores as the salt concentration varies by over six orders of magnitude from $10^{-6}$ mol $L^{-1}$ to 2.5 mol $L^{-1}$, demonstrating a complete breakdown of the linear conductance-concentration scaling law. In dramatic contrast, for nanopores in monolayer $MoS_2$ and $MoSe_2$ which are symmetric in the cross-plane direction, linear scaling holds well at relatively high ion concentrations above $10^{-3}$ mol $L^{-1}$. Atomistic simulations reveal the critical role of dipole-induced asymmetric rearrangement of water molecules confined in the polar nanopores, which leads to higher dielectric barriers that dominate ion transport especially when the pore diameter is around 1 nm.

## II. RESULTS AND DISCUSSION

Controlled fabrication of nanometer-sized pores in monolayer MoSSe is a prerequisite to investigating the effect of the dipole potential on ion transport. Here, nanopores were precisely created using a focused electron beam by employing aberration-corrected scanning transmission electron microscopy (AC-STEM). Compared with traditional methods such as chemical etching [16], ion bombardment [17], and plasma etching [18], AC-STEM-based drilling not only enables angstrom-scale size control but also allows for *in situ* monitoring of the pore formation process [19,20]. To begin with, we grew monolayer $MoS_2$ via chemical vapor deposition (CVD) and then synthesized MoSSe via selenization by replacing the top S layer with Se [21]. Next, the MoSSe sheet was transferred onto a 50 nm-thick $SiN_x$ membrane and suspended across a round aperture of ~200 nm in diameter pre-drilled by a focused $Ga^+$ beam (see Methods, Supplemental Note 1, and Figs. 1-3 in Supplemental Material for details [22]). Subsequently, we employed AC-STEM with an acceleration voltage of 60 kV and an electron beam diameter of ~80 pm to open a single nanopore in the MoSSe sheet by irradiating across a selected region of about 1 nm × 1 nm, which completes the fabrication of a nanopore device. The size and geometry of the pore can be fine-tuned via the irradiation time and area. In Fig. 1(c), we show the STEM image of a representative MoSSe nanopore. To systematically explore and understand the role of the intrinsic dipole, symmetric $MoS_2$ and $MoSe_2$ nanopores were also fabricated using the same method [Fig. 1(c)] for direct comparison.

Prior to device fabrication, the monolayer $MoS_2$, $MoSe_2$, and MoSSe sheets were comprehensively characterized, as shown in Fig. 1(d)-(f). First, Raman spectroscopy revealed that the characteristic out-of-plane and in-plane phonon peaks red-shifted from 405 $cm^{-1}$ ($A_{1g}$) and 384 $cm^{-1}$ ($E_{2g}$) for $MoS_2$ to 290 $cm^{-1}$ and 358 $cm^{-1}$ for MoSSe, respectively [Fig. 1(d)]. These values agree well with previous reports and indicate the successful replacement of the top S atoms with heavier Se atoms [21,64,65]. As for $MoSe_2$, the $A_{1g}$ and $E_{2g}$ peaks further shifted to 241 $cm^{-1}$ and 287 $cm^{-1}$, respectively. Moreover, photoluminescence (PL) spectroscopy also confirmed the conversion [Fig. 1(e)] with the peak energy changing from 1.83 eV ($MoS_2$) to 1.72 eV (MoSSe) and 1.55 eV ($MoSe_2$) [21,64,65]. Finally, atomic-resolution STEM imaging of MoSSe [Fig. 1(f)] clearly identified the positions of the S, Mo, and Se atoms, based on the dependence of the annular dark-field imaging (HAADF) contrast in the intensity profile on the atomic number ($Z$) [66].

To probe ion transport through the TMD nanopores, the devices were each mounted between two reservoirs filled with the same aqueous solutions of salt such as potassium chloride (KCl). A pair of silver/silver chloride (Ag/AgCl) electrodes were then used to apply a bias voltage ($V$) across the two reservoirs and record the corresponding ionic current ($I$) [Fig. 1(b) and Supplemental Material Fig. S4 [22]]. To avoid possible electrochemical etching of the TMD sheets, the bias was limited to within $\pm 0.5$ V [16,67]. To begin with, a device with only the large $SiN_x$ aperture and no TMD coverage was measured in 0.1 M KCl solution as a control, which yielded a linear $I$–$V$ relation (Supplemental Material Fig. S5 [22]) in full agreement with conventional wisdom [68,69] and a large ionic conductance ($G$) of 1390 nS based on Ohm's law. In comparison, measurement of another device with the $SiN_x$ aperture covered by a pristine MoSSe sheet (no nanopore) observed electrical current that was over four orders of magnitude lower from submicro- to pico-ampere (Supplemental Material Fig. S5 [22]), and a much smaller background $G$ of 0.023 nS. This suggested the absence of intrinsic pores or transfer-induced cracks and laid the foundation for subsequently characterizing the TMD nanopore devices.

Before delving into the asymmetric MoSSe nanopores, we first measured the $I$–$V$ characteristics of a symmetric $MoS_2$ nanopore with a diameter of 1.2 nm, which provides a reference for the single-nanopore scenario. Here, we employed KCl solutions of concentration $c$ which varied by over six orders of magnitude from $10^{-6}$ M to 2.5 M [Fig. 2(a)-(b)]. As shown in the inset of Fig. 2(a), for small bias voltages from about -0.2 V to 0.2 V, all the *I–V* curves remain linear which allow the ionic conductances to be determined as the slopes as plotted in Fig. 2(b). For 0.1 M KCl, *G* increases by one to two orders of magnitude compared to the control case with pristine MoSSe [Supplemental Material Fig. S5(b) [22]], indicating the successful fabrication of nanopore.

At high KCl concentrations, *G* appears proportional to *c* which is typical for ion conduction in bulk solutions [Fig. 2(b)]. As the concentration decreases to about $10^{-2}$ M, *G* starts to deviate from the linear relation. At even lower *c* around $10^{-3}$ M, saturation of *G* is observed. This is because the thickness of the electric double layer (EDL) increases notably with decreasing ion concentration [68,70,71]. For example, the EDL expands from ~0.1 nm to ~10 nm as *c* drops from 1 M to $10^{-3}$ M [68]. As a result, the whole nanopore becomes occupied by the EDL, and therefore the ionic transport is dominated by the surface charge and only weakly depends on the ion concentration. Such *G*-*c* relation is commonly observed for ionic transport in various nanochannels [10,72,73].

Subsequently, we proceeded to investigating the ionic response of a MoSSe nanopore with a similar size of 1.1 nm [Fig. 2(c)-(d)]. Following pore fabrication, over 4-fold conductance enhancement was measured in 0.1 M KCl, comparing Fig. 2(d) and Supplemental Material Fig. S5(b) [22]. Intriguingly, we observe a rather distinct and anomalous relation between the ionic conductance and concentration. For *c* below $10^{-3}$ M, *G* exhibits saturation similar to the case of surface charge-governed transport in $MoS_2$ nanopore. However, further increase in *c* does not transform such saturation into a linear scaling, even up to 1 M [Fig. 2(d)]. Instead, the conductance remains essentially flat across the entire range of ion concentration. For more quantitative comparison, the scaling behavior between *G* and *c* can be described by the power-law relation $G \propto c^{\alpha}$. Here, the exponent $\alpha$ is calculated to be nearly 0 for the MoSSe nanopore over the whole concentration range, in dramatic contrast to the $MoS_2$ nanopore where $\alpha$ varies from 0 to 1 with increasing *c*. In addition to KCl, this invariance of the ionic conductance in MoSSe nanopores was also observed in $MgCl_2$ and $LaCl_3$ solutions which feature metal ions of different valences. Moreover, repeated measurements were conducted using a total of 7 different devices of similar pore sizes (~1 nm) and regardless of whether the substitution of Se was performed before or after sample transfer to the $SiN_x$ aperture (Supplemental Material Figs. S6, S7, and Table S1 [22]).

After MoSSe, we further fabricated a $MoSe_2$ nanopore with a diameter of 1.2 nm and measured the ionic conductance in KCl solutions, which exhibits a trend similar to $MoS_2$ [Fig. 2(e)-(f), see Supplemental Note 2 in Supplemental Material for an explanation of the slight difference at low concentrations [22]]. This suggests that the invariant conductance in the MoSSe nanopores is not caused by the existence of Se itself. Intuitively, the intrinsic dipole due to the asymmetric distribution of the S and Se atoms on the two ends of the nanopore must play an essential role. We note that despite the asymmetric structure of the MoSSe nanopores, no discernible rectification was observed (Supplemental Material Figs. S11, S12, S27 to S30, and Supplemental Note 3 [22]), which is partially attributed to the thermodynamically stable Mo-terminated zigzag edge structure formed during STEM etching [48].

To gain further insights into the invariant ionic conductance, we conducted systematic transport measurements in KCl solutions for TMD nanopores with varying sizes. As shown in Fig. 3(a) and (c) and Supplemental Material Figs. S8 and S9 [22], for $MoS_2$ and $MoSe_2$ nanopores with diameters from about 1.0 nm up to 25 nm, $G$ consistently saturates at low concentrations and gradually transitions into a linear increase with $c$ beyond $10^{-3} – 10^{-2}$ M. In comparison, a clear size effect is observed for MoSSe nanopores [Fig. 3(b) and Supplemental Material Fig. S10 [22]]. With a pore diameter of 1.0 nm, $G$ remains practically constant with $c$. As the pore size increases to 2.1 nm, the invariance of $G$ becomes less pronounced above $10^{-2}$ M. At an even larger pore size of 4.1 nm, the $G$-$c$ scaling resembles that observed in $MoS_2$ and $MoSe_2$ nanopores. To quantify trends, we fitted the scaling exponent $\alpha$. As shown in Fig. 3(d), $\alpha \approx 1$ holds for all the $MoS_2$ and $MoSe_2$ nanopores. In contrast, MoSSe nanopores exhibit $\alpha \approx 0$ at diameters around 1.0 nm, which gradually rises to ~1 as the pore size increases to and over 4 nm. Together, these observations highlight the fundamental role of the material selection and pore size for observing invariant conductance at high ion concentrations.

In both biological and artificial systems, variations in the conductance scaling relation have been previously reported, which reflect differences in the way ions permeate through diverse nanopores [70,74-78]. However, to date invariant conductance with ion concentrations spanning six orders of magnitude has never been observed or predicted, and no existing theoretical models can readily elucidate the underlying physical mechanisms.

To help better understand this extraordinary phenomenon, we further measured the ionic conductance at different temperatures from 284.6 K to 302.5 K, using $MoS_2$ and MoSSe nanopores with diameters of about 1.0 nm. Both devices followed the Arrhenius law and yielded activation energies of 20.0 ± 1.7 kJ $mol^{-1}$ and 39.7 ± 4.8 kJ $mol^{-1}$, respectively [Fig. 3(e)]. A higher energy barrier makes it harder for ions to enter the nanopore [36,79], potentially to the extent that increasing the bulk concentration would no longer boost the ion flux through the pore. In general, the energy barrier comes primarily from the surface charge, Coulomb repulsion induced ion-ion interactions, and Born free energy related dehydration effect [55,59,61,80,81]. In Fig. 3(f), we show that the invariant conductance is pH independent which, along with the modest adsorption energy for MoSSe, suggests that surface charge and Coulomb repulsion are unlikely the major factors (see Methods, Figs. S13, S24 to S26, and Supplemental Note 4 in Supplemental Material for details [22]). Therefore, dehydration of ions in confined space is considered the most probable cause of the invariant conductance in MoSSe nanopores.

To see if the ion dehydration process is indeed the key factor, we performed systematic all-atom molecular dynamics (MD) simulations of TMD nanopores in an aqueous solution of KCl at 300 K (see Methods and Fig. S14 in Supplemental Material for details [22]). Due to the negatively charged surface of TMD membranes, as evidenced by zeta potential measurement (Supplemental Material Fig. S13 [22]), cations are the primary species passing through the nanopores [10,82]. Therefore, subsequent simulations primarily focus on the potassium ions ($K^+$). As illustrated in Fig. 4(a), the hydration shell configuration of a $K^+$ ion confined in a nanopore usually differs from that in open space, for example, in terms of the angular distributions of the water molecules around the ion as indicated by the angle $\theta$. To begin with, we considered nanopores with diameters around 1.0 nm. For a $K^+$ ion in bulk solution, there is no preferred position for the surrounding water molecules so the probability distribution of $\theta$ is sinusoidal, which agrees well with previous report [83]. In the nanopores, however, a bimodal distribution is observed with the water molecules located preferably around $\theta = \pi/3$ and $2\pi/3$ [Fig. 4(b)]. Notably, contrary to the symmetric peaks for the $MoS_2$ and $MoSe_2$ nanopores, the MoSSe nanopore displays an asymmetry with a lower peak at $\theta = \pi/3$, which reflects the impact of the intrinsic dipole.

Subsequently, we went on to seek the origin of the distinct hydration structures, hypothesizing that they were a direct consequence of the particular dielectric environment inside the nanopores. To this end, we ran equilibrium MD simulations at 300 K for both bulk water and water confined in nanopores of different diameters from 1.0 nm to 3.8 nm (Supplemental Material Note 5 and Figs. S16-20 [22]). Compared to the bulk, the number density of water molecules in confinement is notably smaller, especially in the MoSSe nanopores around 1.0 nm in diameter (Supplemental Material Figs. S16 and S17 [22]). Moreover, water molecules in these MoSSe nanopores are oriented asymmetrically with a notably broader peak. By employing the Kirkwood-Fröhlich theory [84,85], we further obtained the dielectric constants ($\varepsilon$) (Supplemental Material Figs. S21 and S22 [22]) which determine the energy barriers [Fig. 4(c)] controlling ion translocation through the nanopores according to the Born self-energy theory [86] (Supplemental Material Notes 6 [22]).

Our calculated $\varepsilon$ for bulk water is consistent with the literature value of around 80 [50]. In dramatic contrast, in $MoS_2$, $MoSe_2$, and MoSSe nanopores of 1.0 nm in diameter, $\varepsilon$ drops to 15.2, 21.6, and 13.1, respectively. Correspondingly, the dielectric barriers facing the ions moving from the bulk into the nanopores are 7.2 $k_BT$, 4.6 $k_BT$, and 8.7 $k_BT$. With higher barriers, ion transport becomes increasingly more limited by the entry process. Therefore, under a sufficiently high dielectric barrier, increasing the ion concentration will not enhance the ion flux, which explains the invariant conductance. As the pore diameter increases, the dielectric constants gradually increase which lowers the bulk-pore dielectric barrier [Fig. 4(c) and Supplemental Material Figs. S17-22 [22]]. Meanwhile, the difference between the MoSSe, $MoS_2$, and $MoSe_2$ nanopores becomes smaller and eventually vanishes, which therefore validates our experimental observation of invariant conductance only for MoSSe nanopores around 1.0 nm in diameter. These results highlight the pivotal role of the dielectric environment in ion transport.

Finally, we directly simulated electrically driven ion transport through TMD nanopores of 1.0 nm in diameter to extract the ionic conductance (see Methods and Supplemental Material Fig. S23 for details [22]). Here, we focused on high ion concentrations from 0.5 M to 2.5 M for two primary reasons. First, while invariant conductance at low concentrations is well-documented [10,72,73], experimental observation at high concentrations has never been reported, which stands as the most prominent difference between MoSSe and $MoS_2$/$MoSe_2$ nanopores. Second, these elevated concentrations provide the robust current statistics necessary to overcome the inherent time and length scale limitations of molecular dynamics simulations. As shown in Fig. 4(d), the computed results successfully capture the experimental $G$-$c$ scaling, with an essentially constant conductance for the MoSSe nanopore but increasing conductances for the $MoS_2$ and $MoSe_2$ nanopores.

## III. CONCLUSIONS

In conclusion, we have discovered an invariant ionic conductance in an atomically thin polar nanopore independent of the ion concentration even as it varies by over six orders of magnitude. This has never been reported to the best of our knowledge and is reminiscent of the current saturation in cell membranes at high concentrations. MD simulations reveal that the invariant conductance arises from the substantial bulk-pore energy barrier facing ions entering the MoSSe nanopore, where the dipole-modulated dielectric environment plays a key role. Our results illuminate how confined dipole could be harnessed to construct artificial channels, and represent a novel avenue to regulate ionic transport at angstrom scale beyond existing platforms.

**ACKNOWLEDGMENTS**

This work was funded by the National Natural Science Foundation of China (Grant Nos. 62274004, 52521007, 92577201, 52025023, and T24B2001), the Scientific Research Innovation Capability Support Project for Young Faculty (ZYGXQNJSKYCXNLZCXM-E1) from the Ministry of Education of China, and the National Key R&D Program of China (Grant Nos. 2022YFA1203100, 2023YFB3211200, and 2024YFA1207900). B.S. and K.L. acknowledge support from the New Cornerstone Science Foundation through the XPLORER PRIZE. We thank J. Xu and Z. Liu for the help in preparing $SiN_x$ nanopore by FIB; the Electron Microscopy Laboratory at Peking University for the use of Cs-corrected Nion U-HERMES200 scanning transmission electron microscopy; the Peking Nanofab for fabrication of the $SiN_x$ aperture; and the High-performance Computing Platform of Peking University for supporting the MD and DFT calculations; and the Texas Advanced Computing Center (TACC) at the University of Texas at Austin for the use of the parallel computing resource Lonestar6.

**AUTHOR CONTRIBUTIONS**

L.W. and B.S. conceived the idea and supervised the project. G.X. synthesized the TMD samples under the supervision of K.L.. G.X., N.W., and S.Z. transferred and characterized the samples. S.Z. and J.T. conducted STEM etching of nanopores. S.Z., N.W., J.T., and Z.Z. conducted the ion transport experiments. H.Z. and X.L. performed the classical MD simulations. A.S. and N.A. performed the AIMD simulations. H.Z., J.Y., and N.L. performed the DFT calculations. L.W., B.S., N.A., S.Z., H.Z., N.W., X.L., N.L., A.S., K.L., and G.X. analyzed and discussed the results. B.S., L.W., N.A., S.Z., H.Z., N.W., X.L., A.S., and G.X. wrote the manuscript. S.Z., H.Z., N.W., and G.X. contributed equally. All authors contributed to the review and editing of the manuscript.

**DATA AVAILABILITY**

The data that support the findings of this article are not publicly available. The data are available from the authors upon reasonable request.

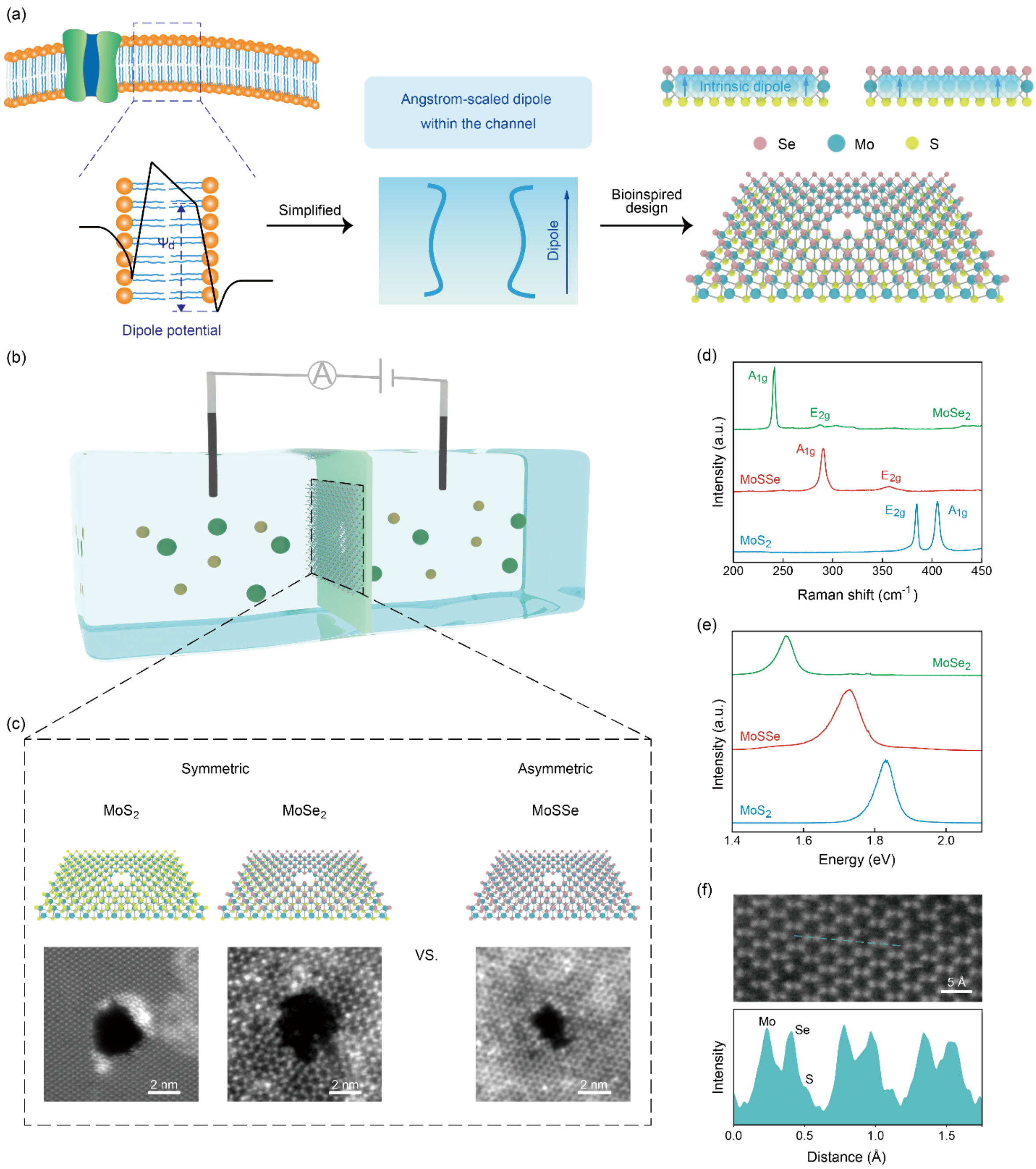

FIG. 1. Nanofluidic devices featuring single polar and nonpolar TMD nanopores. (a) Bioinspired design of a polar MoSSe nanopore to mimic the dipole potential in cell membrane. (b) Schematic of voltage-driven ionic transport through a transition metal dichalcogenide nanopore. (c) Schematic illustrations and AC-STEM images of symmetric nanopores formed in $MoS_2$ and $MoSe_2$, and an asymmetric MoSSe nanopore. (d,e) Raman spectra (d) and photoluminescence spectra (e) demonstrating the successful synthesis of $MoS_2$, $MoSe_2$, and MoSSe. (f) STEM image of a MoSSe sample (top) and an intensity line profile (bottom) showing the location of the Mo, Se, and S atoms.

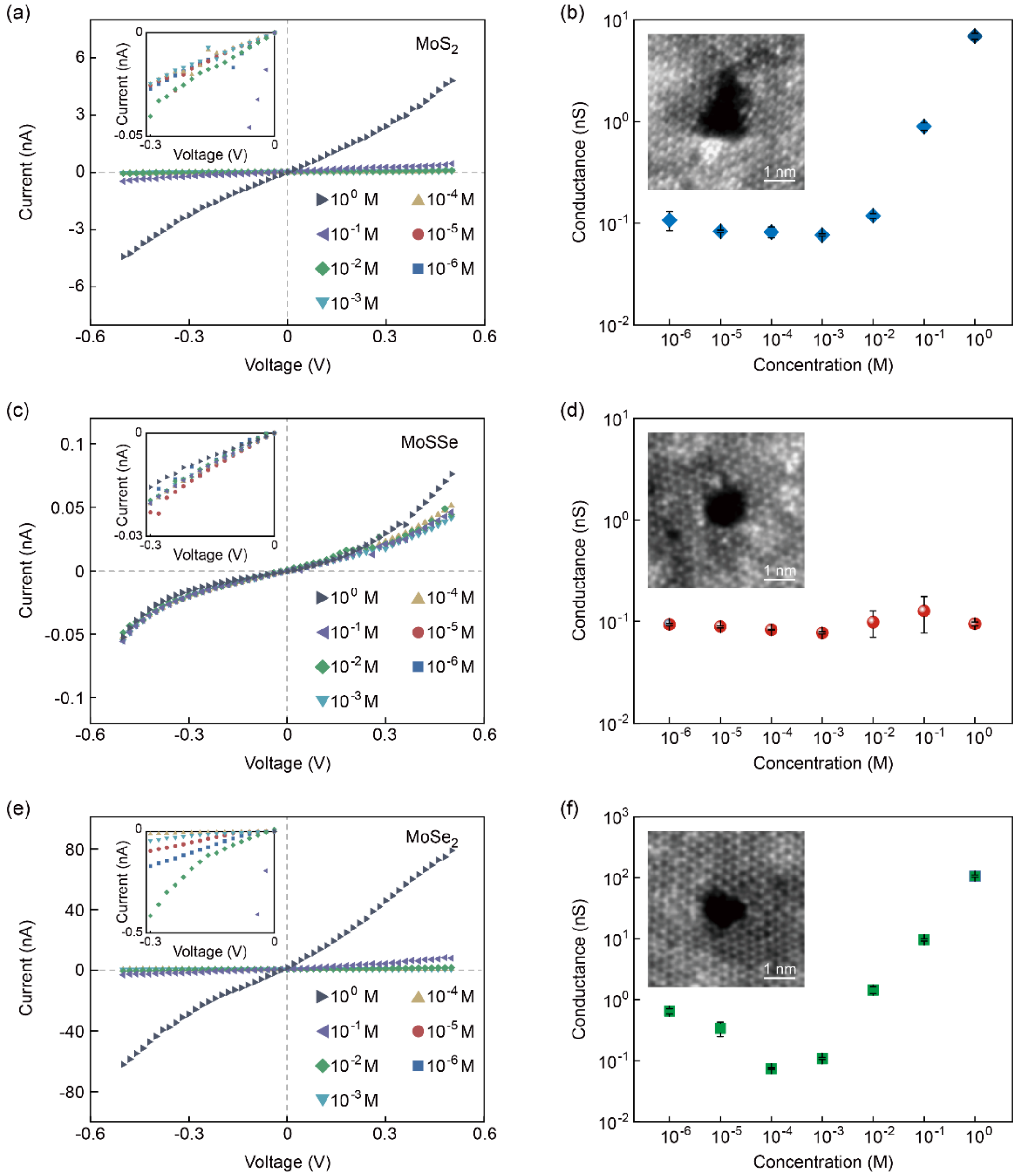

FIG. 2. Invariant conductance in a polar MoSSe nanopore. a,c,e, Measured *I*-*V* curves at different KCl concentrations across $MoS_2$ (a), MoSSe (c), and $MoSe_2$ (e) nanopores. (b,d,f) Ionic conductance extracted from (a), (c), and (e) to show the invariant conductance in the MoSSe nanopore, in contrast to the $MoS_2$ and $MoSe_2$ nanopores. Insets show the AC-STEM images of the $MoS_2$, MoSSe, and $MoSe_2$ nanopores with the effective diameters of 1.2 nm, 1.0 nm, and 1.2 nm, respectively. For the MoSSe nanopore, positive bias means higher voltage on the Se side.

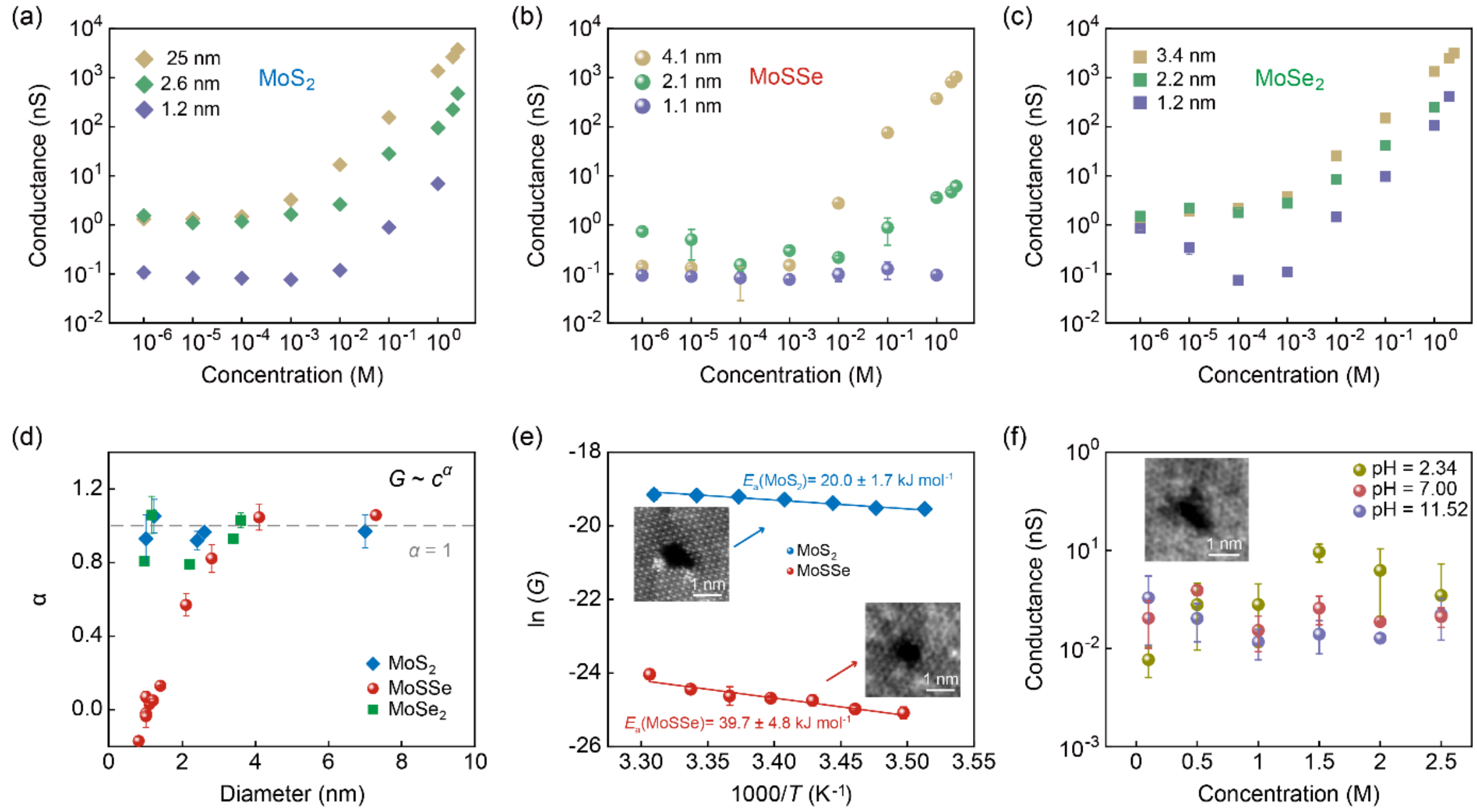

FIG. 3. Identifying the key factor dominating invariant conductance. (a-c) Ionic conductance of $MoS_2$, $MoSe_2$, and MoSSe nanopores with different diameters as a function of KCl concentration. Results for the ~1.0 nm nanopores are taken from FIG. 2 to facilitate direct comparison. (d) The fitted exponent $\alpha$ shows the evolution of invariant conductance in MoSSe nanopores as a function of pore diameter, in contrast to $MoS_2$ and $MoSe_2$ nanopores in which $\alpha$ remains close to 1 regardless of pore size. (e) Ionic conductivity as a function of temperature in 1 nm MoSSe and $MoS_2$ nanopores, together with the activation energies extracted via the Arrhenius equation. Insets show the AC-STEM images. (f) The conductance of a 1 nm MoSSe nanopore under various pH values. Insets show the AC-STEM image.

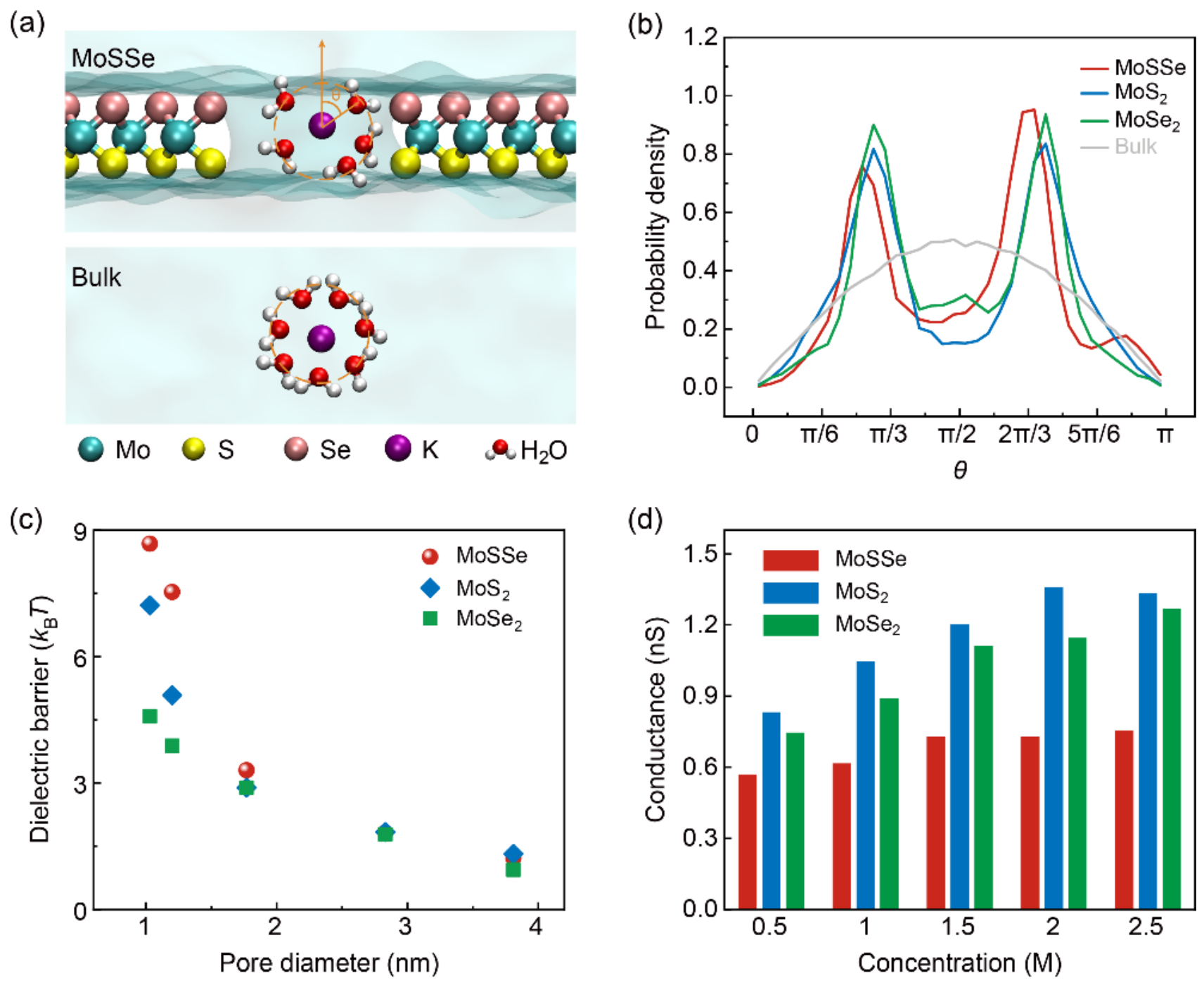

FIG. 4. Mechanism of invariant conductance in polar nanopores. (a) Schematic illustrations of hydrated $K^+$ ions in the MoSSe nanopore (top) and in the bulk KCl solution (bottom), showing the local arrangements of water molecules around the ions. The dashed orange circles represent the hydration shells. $\theta$ is defined as the angle between the vector connecting an ion to the oxygen atom in the first hydration shell and the $z$ axis. (b) The distribution of angular orientations of the water molecules around $K^+$ ions in the $MoS_2$, MoSSe, and $MoSe_2$ nanopores, and also in the bulk solution. (c) Dielectric barriers for $K^+$ ions passing through $MoS_2$, MoSSe, and $MoSe_2$ nanopores with different diameters. (d) Simulated ionic conductance of $MoS_2$, MoSSe, and $MoSe_2$ nanopores for KCl solutions with the concentration varying from 0.5 M to 2.5 M.